\documentclass[12pt]{article}
\usepackage{geometry}
\usepackage{amssymb}
\usepackage{amsmath}
\usepackage{bbm}
\usepackage{MnSymbol}
\usepackage{hyperref}
\usepackage{braket}
\usepackage{cite}
\usepackage{xcolor}
\usepackage{subcaption,tikz}
\usetikzlibrary{arrows,decorations.markings,decorations.pathreplacing,decorations.shapes,decorations.pathmorphing}

\newcommand{\be}{\begin{equation}}
	\newcommand{\ee}{\end{equation}}
\newcommand{\bea}{\begin{eqnarray}}
	\newcommand{\eea}{\end{eqnarray}}

\newcommand{\lrr}{\longrightarrow}

\newcommand{\Z}{\mathbb{Z}}
\newcommand{\A}{\mathcal{A}}
\newcommand{\id}{\mathbbm{1}}

\newcommand{\D}{\mathbb{D}}

\def\sqr#1#2{{\vcenter{\vbox{\hrule height.#2pt
				\hbox{\vrule width.#2pt height#1pt \kern#1pt
					\vrule width.#2pt}\hrule height.#2pt}}}}

\def\square
{\mathop{\mathchoice{\sqr{12}{15}}{\sqr{9}{12}}
		{\sqr{6.3}{9}}{\sqr{4.5}{9}}}}
	
\def\greysquare{\mathop{\fcolorbox{black}{gray}{\rule{0pt}{6pt}\rule{6pt}{0pt}}}}

\tikzset{
	partial ellipse/.style args={#1:#2:#3}{
		insert path={+ (#1:#3) arc (#1:#2:#3)}
	}
}

\numberwithin{equation}{section}

\begin{document}
	
\begin{flushright}
	MI-HET-816
\end{flushright}

\begin{center}
	
	{\large\bf Gauge Defects}\\
	or {\large\bf Gauging as Symmetry}\\

	\vspace*{0.2in}
	
	Thomas Vandermeulen
	
	\vspace*{0.2in}
	
	{\begin{tabular}{l}
			George P.~and Cynthia W.~Mitchell Institute\\
			for Fundamental Physics and Astronomy\\
			Texas A\&M University\\
			College Station, TX 77843 \end{tabular}}
	
	\vspace*{0.2in}
	
	{\tt tvand@tamu.edu}
	
\end{center}
\pagenumbering{gobble}
\baselineskip=18pt

Gauging a symmetry can be thought of as the insertion of a spacetime-filling defect.  Accordingly, we regard each gaugeable symmetry in a theory as defining a $-1$-form symmetry via condensation.  The resulting operators, called gauge defects, have a natural fusion product, generally non-invertible, which we explore in a variety of two-dimensional theories.

\newpage
\tableofcontents

\newpage

\section{Introduction}
\label{sec:intro}
\pagenumbering{arabic}

In the modern approach, one regards the existence of a symmetry as equivalent to the existence of the topological operators that implement that symmetry \cite{GKSW}.  Gauging that symmetry is then understood as flooding the spacetime with a network of its topological defects, which must form a mesh which is sufficiently fine to act on all operators present a given correlation function.

There exists a related procedure by which one constructs a new topological defect from an existing symmetry \cite{Gaiotto:2019xmp,RSS,Choi:2022zal}.  Specifically, we can take any higher-form symmetry in a theory and construct a topological operator by flooding a subsurface of the spacetime with that defect.  For instance, a commonly-studied application is the condensation of topological defect lines (TDLs) on a 2d surface in three spacetime dimensions, building a 0-form symmetry operator out of a 1-form symmetry.  One can then study the fusion of the resulting surfaces, and one finds that they form higher representation categories of the original symmetry \cite{UNI,Bartsch:2022mpm,Bhardwaj:2022kot,Bartsch:2022ytj,Delcamp:2023kew,Bhardwaj:2023wzd,Bartsch:2023pzl,Bhardwaj:2023ayw}.\footnote{Mathematically, gaugings correspond to module categories over the symmetry being gauged \cite{BhardwajTachikawa}.  The fusion of these module categories is then governed by the Deligne tensor product \cite{etingof2009fusion}.}

Nothing prevents us from extending the above description to the whole of spacetime.  In this case, the condensation defect constitutes a $-1$-form symmetry.  We will thus interpret the ordinary gauging (0-gauging, in the condensation defect language) of any symmetry as the insertion of a spacetime-filling condensation defect which describes a $-1$-form symmetry of the theory.  Said another way, we will argue that there exists a $-1$-form symmetry associated with any gaugeable (possibly higher-form) symmetry in a theory.  The resulting topological defect, which we will call a gauge defect, is exactly the network of lower-dimensional defects referred to in the conventional description of gauging.

\section{Warmup: Gauge Defects in 1d Theories}

As a first example of gauge defects, we will examine (0+1)d (quantum mechanical) theories with symmetry.  Recall that 0-form symmetries in quantum mechanical theories correspond to topological local operators.  Due to the presence of these non-trivial topological local operators, a quantum mechanical theory with symmetry has multiple ground states.  Given a 0-form symmetry $G$, let the corresponding local operators be $\sigma_g$, which fuse according to the group law:
\be
\sigma_{g_1}\otimes\sigma_{g_2}=\sigma_{g_1g_2}.
\ee
We can make the decomposition of the theory into multiple disjoint theories with unique ground states manifest via a change of basis.  Given an irreducible representation $R$ of $G$, write
\be
\pi_R = \frac{\text{dim}(R)}{|G|}\sum_{g\in G}\chi_R(g)\sigma_g.
\ee
These $\pi$ are orthonormal projectors; that is, their fusion is
\be
\pi_R\otimes\pi_{R'} = \delta_{R,R'}\pi_R.
\ee
Physically, the theory breaks into a direct sum of subtheories with unique groundstate/vacuum $\pi_R$, also referred to as universes.

\begin{figure}
	\centering
	\begin{tikzpicture}
		\draw[decorate,decoration={crosses,segment length=3pt}] (0,0) circle (1cm);
		\node at (0,1.3) {$\D(\Z_2)_{\pm}$};
		\node[scale=1.5] at (1.5,0) {$=$};
		\node[scale=1.5] at (2.1,0) {$\frac{1}{2}$};
		\draw[thick] (2.5,-1) -- (2.5,1);
		\draw[thick] (2.5,1) -- (2.75,1);
		\draw[thick] (2.5,-1) -- (2.75,-1);
		\draw (4,0) circle (1cm);
		\filldraw[black] (4,1) circle (3pt);
		\node[scale=1.5] at (4,1.4) {$\sigma_1$};
		\node[scale=1.5] at (5.5,0) {$\pm$};
		\draw (7,0) circle (1cm);
		\filldraw[black] (7,1) circle (3pt);
		\node[scale=1.5] at (7,1.4) {$\sigma_g$};
		\draw (8.5,-1) -- (8.5,1);
		\draw (8.5,1) -- (8.25,1);
		\draw (8.5,-1) -- (8.25,-1);
	\end{tikzpicture}
	\caption{The $\D(\Z_2)_\pm$ gauge defect on an empty worldcircle.}
	\label{z2_cond_1d}
\end{figure}

Gauging in a 1d theory corresponds to projecting onto a subset of universes.  This is accomplished by flooding the spacetime with (possibly a sum of) projectors corresponding to one (or more) of the irreducible representations of $G$.  The choice of $R$ is the discrete torsion for the gauging, classified by $H^1(G,U(1))$.  Simple gauge defects (that is, those that cannot be expressed as a sum) are thus labeled by irreducible representations.  These are line operators formed by condensations of projectors (which are the condensation defect projectors described in \cite{LRS}) -- we will write the gauge defect obtained by condensing $\pi_R$ as $\D(G)_R$.

We can deduce the fusion rules for these defects by examining their action on an otherwise empty worldsheet.  In this case, we are able to collapse the condensation defect down to a single projector, which can further be rewritten in terms of the symmetry defects $\sigma_g$.  This is shown for a $\Z_2$ symmetry in Figure~\ref{z2_cond_1d}, where an insertion of the gauge defect $\D(\Z_2)_{\pm}$ (where $\pm$ labels the $\Z_2$ irreps) on an empty, compact worldline is equivalent to the insertion of $\pi_{\pm}=\frac{1}{2}(\sigma_1\pm\sigma_g)$.  Inserting multiple gauge defects is then equivalent to inserting multiple projectors, so the gauge defects inherit the orthonormal fusion relations of the $\pi$:
\be
\D(G)_R\otimes\D(G)_{R'} = \delta_{R,R'}\D(G)_R.
\ee
Note that the identity background defect is not a simple gauge defect -- it is obtained by summing the $\pi_R$ over irreps $R$.

There is another notable source of $-1$-form symmetries in 1d theories, which is quantum symmetries: the dual to a 0-form symmetry in 1d is a $-1$-form symmetry.  These quantum symmetries are examples of background (i.e.~spacetime-filling) defects which are not gauge defects, as the gauged theories need not have any 0-form symmetry operators to condense.  As their action on a theory is to produce additional copies of that theory, they generically fuse non-invertibly \cite{LFS}.

\section{Gauge Defects in 2d Theories}

Now we move up a dimension to examine generic (1+1)-dimensional theories, where the same techniques as before will lead us to condensation defects with less trivial fusion relations.  As 2d theories can contain both 0-form and 1-form symmetries, we will see the effect of mixing gauge defects from both types.

\subsection{Non-Anomalous, Abelian 0-Form Symmetries}

\begin{figure}
	\begin{tikzpicture}
		\draw[thin] (0.5,0)--(2.5,0);
		\draw[thin] (2.5,0)--(2.5,2);
		\draw[thin] (0.5,0)--(0.5,2);
		\draw[thin] (0.5,2)--(2.5,2);
		\fill[gray!20,opacity=0.5] (0.5,0) -- (0.5,2) -- (2.5,2) -- (2.5,0) -- cycle;
		\node[scale=1.5] at (1.5,1) {$\D(\Z_2)$};
		\node[scale=1.5] at (3.2,1) {$=\frac{1}{2}$};
		\draw[thick] (3.75,-0.25) -- (3.75,2.25);
		\draw[thick] (3.75,-0.25) -- (4,-0.25);
		\draw[thick] (3.75,2.25) -- (4,2.25);
		\draw[thin] (4,0)--(6,0);
		\draw[thin] (6,0)--(6,2);
		\draw[thin] (4,0)--(4,2);
		\draw[thin] (4,2)--(6,2);
		\node[scale=1.5] at (6.5,1) {$+$};
		\draw[thin] (7,0)--(9,0);
		\draw[thin] (9,0)--(9,2);
		\draw[thin] (7,0)--(7,2);
		\draw[thin] (7,2)--(9,2);
		\draw[thick,->] (8,0) -- (8,1);
		\draw[thick] (8,1) -- (8,2);
		\node at (8.4,1.4) {$g$};
		\node[scale=1.5] at (9.5,1) {$+$};
		\draw[thin] (10,0)--(12,0);
		\draw[thin] (12,0)--(12,2);
		\draw[thin] (10,0)--(10,2);
		\draw[thin] (10,2)--(12,2);
		\draw[thick,->] (10,1)--(11,1);
		\draw[thick] (11,1)--(12,1);
		\node at (10.6,1.4) {$g$};
		\node[scale=1.5] at (12.5,1) {$+$};
		\draw[thin] (13,0)--(15,0);
		\draw[thin] (15,0)--(15,2);
		\draw[thin] (13,0)--(13,2);
		\draw[thin] (13,2)--(15,2);
		\draw[thick,->] (14,0)--(14,1.5);
		\draw[thick] (14,1.5)--(14,2);
		\draw[thick,->] (13,1)--(13.5,1);
		\draw[thick] (13.5,1)--(15,1);
		\node at (14.4,1.6) {$g$};
		\node at (13.4,0.6) {$g$};
		\draw[thick] (15.25,-0.25) -- (15.25,2.25);
		\draw[thick] (15.25,-0.25) -- (15,-0.25);
		\draw[thick] (15.25,2.25) -- (15,2.25);
	\end{tikzpicture}
	\caption{The $\D(\Z_2)$ gauge defect on an empty toroidal worldsheet.}
	\label{z2_cond_2d}
\end{figure}

Again we begin with the simplest example, a $\Z_2$ 0-form symmetry.  There are two gauge defects: the trivial one $\D(\id)$ and a non-trivial one $\D(\Z_2)$ which comes from condensing the object $1+g$ (and corresponds to gauging the $\Z_2$).  To determine the fusion of $\D(\Z_2)$ with itself, we can put it on an empty torus.  As its insertion corresponds to gauging $\Z_2$, this must give us the $\Z_2$ orbifold partition function.  Thus, we expect to have

\be
\label{dz2}
\greysquare^{\D(\Z_2)}=\frac{1}{2}\left[{\scriptstyle 1}\square_1+{\scriptstyle 1}\square_g+{\scriptstyle g}\square_1+{\scriptstyle g}\square_g\right].
\ee
This equation is depicted in more detail in Figure~\ref{z2_cond_2d}.  In order to compute $\D(\Z)2\otimes\D(\Z_2)$, we insert $D(\Z_2)$ into the right-hand side of (\ref{dz2}) and apply the equation iteratively.\footnote{One could rightfully raise objections to this approach, as (\ref{dz2}) was only meant to hold on an otherwise empty worldsheet.  For simple examples like this one, however, we will get away with it.  Subsequent sections will include examples where this is not the case.}  Explicitly, this results in
\begin{multline}
\label{dz2dz2}
\greysquare^{\D(\Z_2)\otimes\D(\Z_2)}=\frac{1}{2}\left[{\scriptstyle 1}\greysquare_1^{\D(\Z_2)}+{\scriptstyle 1}\greysquare_g^{\D(\Z_2)}+{\scriptstyle g}\greysquare_1^{\D(\Z_2)}+{\scriptstyle g}\greysquare_g^{\D(\Z_2)}\right]\\
= \frac{1}{2}\bigg[\frac{1}{2}\left({\scriptstyle 1\otimes1}\square_{1\otimes1}+{\scriptstyle 1\otimes 1}\square_{1\otimes g}+{\scriptstyle 1\otimes g}\square_{1\otimes1}+{\scriptstyle 1\otimes g}\square_{1\otimes g}\right)+\frac{1}{2}\left({\scriptstyle 1\otimes 1}\square_{g\otimes 1}+{\scriptstyle 1\otimes 1}\square_{g\otimes g}+{\scriptstyle 1\otimes g}\square_{g\otimes 1}+{\scriptstyle 1\otimes g}\square_{g\otimes g}\right)\\
+\frac{1}{2}\left({\scriptstyle g\otimes 1}\square_{1\otimes 1}+{\scriptstyle g\otimes 1}\square_{1\otimes g}+{\scriptstyle g\otimes g}\square_{1\otimes 1}+{\scriptstyle g\otimes g}\square_{1\otimes g}\right)+\frac{1}{2}\left({\scriptstyle g\otimes 1}\square_{g\otimes 1}+{\scriptstyle g\otimes 1}\square_{g\otimes g}+{\scriptstyle g\otimes g}\square_{g\otimes 1}+{\scriptstyle g\otimes g}\square_{g\otimes g}\right)\bigg]\\
= {\scriptstyle 1}\square_1+{\scriptstyle 1}\square_g+{\scriptstyle g}\square_1+{\scriptstyle g}\square_g = \greysquare^{\D(\Z_2)}+\greysquare^{\D(\Z_2)},
\end{multline}
which tells us that $D(\Z_2)\otimes\D(\Z_2)=2D(\Z_2)$.  We can summarize the fusion of gauge defects in a $\Z_2$-symmetric 2d theory as
\be
\begin{tabular}{r c | c c}
	& $\otimes$ & 1 & $\alpha$ \\\cline{2-4}
	$\D(\id)=$ & 1 & 1 & $\alpha$ \\
	$\D(\Z_2)=$ & $\alpha$ & $\alpha$ & $2\alpha$
\end{tabular}.
\ee
These are the fusion rules of the higher representation category 2Rep$(\Z_2)$, which are identical to the $\Z_2$ Burnside ring \cite{douglas2018fusion}.

A few comments are in order.  First, note that the result here is less trivial than the 1d case where we had orthonormal fusion of the gauge defects.  Even for the simplest example of a $\Z_2$ symmetry, the gauge defects fuse non-invertibly.  Also it is worth pointing out that iterating the $\Z_2$ gauge defect is not the same thing as gauging $\Z_2$ then gauging the Rep$(\Z_2)$ quantum symmetry of the gauged theory.  While successive gaugings of a theory can be composed to produce a gauging of the original theory, the defects treated here all correspond to gaugings of a single theory.  We can understand this better by rearranging (\ref{dz2dz2}) to get
\be
\frac{1}{2}\left[{\scriptstyle 1}\greysquare_1^{\D(\Z_2)}+{\scriptstyle 1}\greysquare_g^{\D(\Z_2)}+{\scriptstyle g}\greysquare_1^{\D(\Z_2)}+{\scriptstyle g}\greysquare_g^{\D(\Z_2)}\right] = \frac{1}{2}\left[{\scriptstyle 1}\greysquare_1^{\D(\Z_2)}+{\scriptstyle 1}\greysquare_1^{\D(\Z_2)}+{\scriptstyle 1}\greysquare_1^{\D(\Z_2)}+{\scriptstyle 1}\greysquare_1^{\D(\Z_2)}\right].
\ee
That is, our $\Z_2$ symmetry acts trivially in the $\D(\Z_2)$ background.  Of course, this is exactly what happens when we gauge a symmetry -- intuitively, we are quotienting the symmetry out of the theory, causing it to act trivially in the gauged theory.  It should be no surprise, then, that successive gauging of the same symmetry produces the partition function for the pure gauge theory associated with that symmetry.\\

We can repeat this exercise easily enough for a larger group.  Our first example had no discrete torsion, so next we examine $\Z_2\times\Z_2$ -- the smallest group for which $H^2(G,U(1))$ is non-trivial.  We have a simple gauge defect for each subgroup of $\Z_2\times\Z_2$, along with a choice of discrete torsion for that subgroup.  This gives six defects: the trivial one, three corresponding to $\Z_2$ subgroups, and two corresponding to the whole group with the two possible choices of discrete torsion.  Calculations along the lines of (\ref{dz2dz2}) lead to the following fusion rules for these six defects:
\be
\begin{tabular}{r c | c c c c c c}
	& $\otimes$ & 1 & $\alpha$ & $\beta$ & $\gamma$ & $\delta$ & $\epsilon$\\\cline{2-8}
	$\D(\id)=$ & 1 & 1 & $\alpha$ & $\beta$ & $\gamma$ & $\delta$ & $\epsilon$\\
	$\D(\Z_2)=$ & $\alpha$ & $\alpha$ & $2\alpha$ & $\delta$ & $\delta$ & $2\delta$ & $\alpha$\\
	$\D(\Z_2)=$ & $\beta$ & $\beta$ & $\delta$ & $2\beta$ & $\delta$ & $2\delta$ & $\beta$\\
	$\D(\Z_2)=$ & $\gamma$ & $\gamma$ & $\delta$ & $\delta$ & $2\gamma$ & $2\delta$ & $\gamma$\\
	$\D(\Z_2\times\Z_2)_+=$ & $\delta$ & $\delta$ & $2\delta$ & $2\delta$ & $2\delta$ & $4\delta$ & $\delta$ \\
	$\D(\Z_2\times\Z_2)_-=$ & $\epsilon$ & $\epsilon$ & $\alpha$ & $\beta$ & $\gamma$ & $\delta$ & 1
\end{tabular}
\ee
where the subscript on $\D(\Z_2\times\Z_2)_\pm$ indicates the choice of discrete torsion in $H^2(\Z_2\times\Z_2)=\Z_2$.  These are the fusion rules of 2Rep$(\Z_2\times\Z_2)$ \cite{UNI,Bartsch:2022mpm,Decoppet}.

\subsection{Anomalous 0-Form Symmetries}
\label{sec:anomalies}

These results need to be modified in the presence of an 't Hooft anomaly, which would obstruct gauging part of the symmetry.  In the easiest scenario, there will be a unique, maximal non-anomalous subgroup $K$ of our symmetry $G$, and the gauge defects in the $G$-symmetric theory will correspond to subgroups of $K$ along with a choice of discrete torsion for each.  However, this is not the general case; there may be multiple non-trivial non-anomalous subgroups with trivial intersection.

For example, consider a 2d theory with $\Z_2\times\Z_2$ symmetry and a mixed anomaly between the two $\Z_2$ factors -- that is, two of the three $\Z_2$ subgroups are non-anomalous (and the full $\Z_2\times\Z_2$ is therefore also anomalous).  In this case, we expect two non-trivial gauge defects corresponding to the two gaugeable $\Z_2$ subgroups.  Label these subgroups $A$ and $B$.  The self-fusion of $\D(A)$ and $\D(B)$ is easy enough, as it should form 2Rep$(\Z_2)$.  The cross-terms will require slightly more work.

To see the issue, note that calculating $\D(A)\otimes\D(B)$ requires knowing the result of gauging the symmetry $B$ in the theory gauged by $A$.  In the non-anomalous $\Z_2\times\Z_2$ example, this was the same as gauging $A\times B$ in the original theory; the same answer cannot apply here, as $A\times B$ is anomalous and has no gauge defect associated with it.

In fact, the effect of the mixed anomaly is that $B$ will not be (directly) gaugeable in the $A$-gauged theory.  Recall that gauging has the effect of swapping mixed anomalies and extension classes \cite{Tachikawa}, in such a way that the symmetry of the theory gauged by $A$ will be $\hat{A}.B\simeq\Z_4$, where a dot indicates extension.  Thus, in this theory, $B$ is no longer a standalone gaugeable symmetry.  What happens to the $\D(B)$ gauge defect, then?

For a reasonable guess, we can appeal to our intuition that this defect is woven from $B$ lines and their self-fusions.  Letting $b$ generate $B$, in the original theory we had $b\otimes b=\id$, so the gauge defect would in general break into a combination of $b$ lines and identity lines (as in (\ref{dz2dz2})).  In the $A$-gauged theory, however, $b\otimes b$ is no longer trivial -- we instead have $b\otimes b =\hat{a}$ where $\hat{a}$ generates the $\Z_2$ quantum dual to $A$.  So, we see that we cannot have a gauge defect for $B$ lines in this theory without the inclusion of $\hat{A}$ lines; basically, the only way to gauge $B$ is to gauge the full $\hat{A}.B$.  Since $\hat{A}$ is dual to $A$, this has the effect of undoing the original $A$ gauging while also gauging $B$.  Therefore we expect the fusion rules for gauge defects in this $\Z_2\times\Z_2$ theory with mixed anomaly to take the form

\be
	\begin{tabular}{r c | c c c}
		& $\otimes$ & $1$ & $\alpha$ & $\beta$ \\\cline{2-5}
		$\D(\id)=$ & $1$ & $1$ & $\alpha$ & $\beta$ \\
		$\D(A)=$ & $\alpha$ & $\alpha$ & $2\alpha$ & $\beta$ \\
		$\D(B)=$ & $\beta$ & $\beta$ & $\alpha$ & $2\beta$ \\
	\end{tabular}.
\ee

We might well have expected the presence of an anomaly to impact the fusion of the gauge defects, and we see that quite strikingly through the failure of both commutativity and associativity in the result.

\subsection{Non-Invertible Symmetries}
\label{sec:nonab}

Now we would like to consider gauge defects in a theory with non-invertible symmetry.  The simplest example of a gaugeable non-invertible symmetry is the fusion category Rep$(S_3)$ formed by representations of the symmetric group $S_3$.  Rep$(S_3)$ has three simple objects 1, $X$ and $Y$ corresponding respectively to the trivial irrep, the sign irrep, and the two-dimensional irrep of $S_3$.  The Rep$(S_3)$ fusion rules are
\be
	\begin{tabular}{c | c c c}
		$\otimes$ & $1$ & $X$ & $Y$ \\\cline{1-4}
		$1$ & $1$ & $X$ & $Y$ \\
		$X$ & $X$ & $1$ & $Y$ \\
		$Y$ & $Y$ & $Y$ & $1+X+Y$ \\
	\end{tabular}.
\ee

First we will write out gauge defects in a theory with Rep$(S_3)$ symmetry.  The simple gauge defects correspond to Morita equivalence classes of haploid, special symmetric Frobenius algebras in Rep$(S_3)$, and they are in one-to-one correspondence with the gaugeable symmetries of $S_3$.  The algebras dual to gauging $S_3$, $\Z_3$, $\Z_2$ and $\id$ have algebra objects $1$, $1+X$, $1+Y$ and $1+X+2Y$, respectively, and we will refer to these symmetries by their algebra objects.

The $1+X$ algebra is a $\Z_2$ subsymmetry, so its gauge defects should fuse as 2Rep$(\Z_2)$.  $1+X+2Y$ is the regular representation, and gauging it produces a theory with $S_3$ symmetry.  We can interpret the fusion of $\D(1+X+2Y)$ with itself as gauging the regular representation of Rep$(S_3)$ in Rep$(S_3)$ gauge theory, and the fusion coefficient should be the partition function of SPT(Rep$(S_3)$)/$(1+X+2Y)$.  This theory has six ground states which transform under a free action of $S_3$, so we expect $\D(1+X+2Y)\otimes\D(1+X+2Y)=6\D(1+X+2Y)$.

The fusion of $\D(1+Y)$ with itself is more subtle.  So far the coefficients of self-fusion of gauge defects for 0-form symmetries have been the partition function of gauge theory for those symmetries.  There is, however, no SPT for the $1+Y$ symmetry (this is related to the fact that 1 and $Y$ do not form a subsymmetry of Rep$(S_3)$), so we need to put some more thought into how $Y$ acts in the theory gauged by $1+Y$ (which again has Rep$(S_3)$ symmetry).  Normally a gauged symmetry acts trivially in the gauged theory.  But note that in order for $Y$ to act trivially in Rep$(S_3)$, we would need $Y=1+1$ (two copies of the identity), but this is inconsistent with the fusion rules unless $X$ also acts trivially.  However, it would be consistent for $Y$ to act as $1+X$.\footnote{This does, in fact, happen: $Y$ acts as $1+X$ (instead of trivially, as one might expect) in the Rep$(S_3)$-symmetric irreducible TQFT in which the $\Z_2$ symmetry is spontaneously broken \cite[section 5.3.2]{bhardwaj2023gapped}.}  So a reasonable guess is that the $\D(1+Y)$ gauge defect acts as $\D(1)\oplus\D(1+X)$ in the theory gauged by $1+Y$, which suggests the fusion rule $\D(1+Y)\otimes\D(1+Y)=\D(1+Y)\otimes(\D(1)\oplus\D(1+X)) = \D(1+Y)\oplus\D(1+X+2Y)$.  Filling in the cross-terms leads us to

\be
\label{2reps3}
\begin{tabular}{r c | c c c c}
	& $\otimes$ & 1 & $\alpha$ & $\beta$ & $\gamma$ \\\cline{2-6}
	$\D(1)=$ & 1 & 1 & $\alpha$ & $\beta$ & $\gamma$\\
	$\D(1+X)=$ & $\alpha$ & $\alpha$ & $2\alpha$ & $\gamma$ & $2\gamma$\\
	$\D(1+Y)=$ & $\beta$ & $\beta$ & $\gamma$ & $\beta+\gamma$ & $3\gamma$ \\
	$\D(1+X+2Y)=$ & $\gamma$ & $\gamma$ & $2\gamma$ & $3\gamma$ & $6\gamma$
\end{tabular},
\ee
which is indeed the fusion of 2Rep$(S_3)$ \cite[section 3.2.4]{UNI}.  Note additionally that the fusion ring matches the $S_3$ Burnside ring, as expected \cite{Greenough_2010}.  Appendix~\ref{sec:rs3_defects} reproduces this result via the theta defect method of \cite{UNI}, which generates condensation defects in a gauged theory by stacking with topological quantum field theories (TQFTs) before gauging.  This technique also allows us, in Appendix~\ref{sec:s3_defects}, to define gauge defects for an $S_3$-symmetric theory using Rep$(S_3)$-symmetric TQFTs, though doing so requires slightly more care.  While we are able to identify the expected simple defects by this method, we leave issues of their fusion to future work.

\section{2d Theories with Mixed Symmetry}

We can now add in 1-form symmetries to our 2d systems.  These are controlled by topological local operators, which means they will behave like 0-form symmetries in 1d.

\subsection{Disjoint Unions}

As a first example, we can take the direct sum of two $\Z_2$-symmetric theories.  We can regard this disjoint union as having a $\Z_2^1$ symmetry, where superscript $p$ indicates a $p$-form symmetry.  Associated with the 1-form symmetry we have projectors $\pi_\pm$, just as in the 1d case.  These allow us to split the gauge defects among the universes, so in total we find the fusion rules

\be
\begin{tabular}{r c | c c c c}
	& $\otimes$ & $1_+$ & $\alpha_+$ & $1_-$ & $\alpha_-$ \\\cline{2-6}
	$\D(\Z_2^1)_+=$ & $1_+$ & $1_+$ & $\alpha_+$ & 0 & 0\\
	$\D(\Z_2^0)_+=$ & $\alpha_+$ & $\alpha_+$ & $2\alpha_+$ & 0 & 0\\
	$\D(\Z_2^1)_-=$ & $1_-$ & 0 & 0 & $1_-$ & $\alpha_-$ \\
	$\D(\Z_2^0)_-=$ & $\alpha_-$ & 0 & 0 & $\alpha_-$ & $2\alpha_-$
\end{tabular},
\ee
which is of course 2Rep$(\Z_2)\oplus$2Rep$(\Z_2)$, exactly as we would expect from the direct sum of two theories with $\Z_2^0$ symmetry.\footnote{Note that in the case that the two universes are isomorphic, there is an additional $\Z_2^0$ given by exchanging the two copies.  This is the situation in which we have the direct product of a single theory with $\Z_2$ gauge theory; gauge defects from the exchange symmetry in this theory will be examined in section~\ref{sec:tqft}.}  The general expectation is that in $d$-dimensional theories, the presence of a $(d-1)$-form symmetry will lead to a direct sum decomposition of the gauge defect fusion rules.\\

The above extends straightforwardly to theories with 2-group symmetry, which is the case where a 0-form symmetry is non-trivially extended by a 1-form symmetry.  That is, our total symmetry $K^1.G^0$ fits into a short exact sequence
\be
1\lrr K^1\lrr K^1.G^0 \lrr G^0 \lrr 1.
\ee
Sequences of this form are classified by $H^3(G,K)$, so each non-trivial 3-group comes accompanied by a non-trivial class $\beta\in H^3(G,K)$ known as the Postnikov class.

Due to the presence of the unobstructed 1-form symmetry, we should still expect to find a decomposition.  The impact of the Postnikov class will be to determine the 't Hooft anomaly of $G^0$ in each universe.  More specifically, given a universe corresponding to an irrep $R$, the anomaly in the local $G^0$ symmetry is given by
\be
\label{2group_anom}
\omega_R = \beta\cup\chi_R\in H^3(G,U(1)).
\ee
Any subgroup $B$ of $G$ for which $\omega_R$ does not pull back to the trivial class in $H^3(B,U(1))$ will not be gaugeable, and there will therefore not be a gauge defect associated with it.  So the fusion of the gauge defects will once again decompose into a sum over universes, with defects now corresponding to the non-anomalous subgroups of $G$ in each universe.

The simplest example of a 2-group is $\Z_2^1.\Z_2^0$, where we can take the non-trivial class in $H^3(\Z_2,\Z_2)=\Z_2$ as the Postnikov.  There are two universes, which we again label by a plus and minus, corresponding respectively to the trivial and non-trivial $\Z_2$ characters.  From (\ref{2group_anom}) one straightforwardly calculates that the $\Z_2^0$ in the plus universe is non-anomalous, while the $\Z_2^0$ in the minus universe is anomalous.  The latter universe therefore has no gaugeable symmetries, and only the identity gauge defect.  So the fusion rules of gauge defects in a 2d theory with 2-group symmetry $\Z_2^1.\Z_2^0$ are

\be
\begin{tabular}{r c | c c c}
	& $\otimes$ & $1_+$ & $\alpha_+$ & $1_-$ \\\cline{2-5}
	$\D(\Z_2^1)_+=$ & $1_+$ & $1_+$ & $\alpha_+$ & 0 \\
	$\D(\Z_2^0)_+=$ & $\alpha_+$ & $\alpha_+$ & $2\alpha_+$ & 0 \\
	$\D(\Z_2^1)_-=$ & $1_-$ & 0 & 0 & $1_-$ \\
\end{tabular},
\ee
which is 2Rep$(\Z_2)\oplus$2Rep$(\id)$.

\subsection{TQFTs}
\label{sec:tqft}

Topological quantum field theories in two dimensions contain a mix of 0-form and 1-form symmetries, which will ultimately simplify the gauge defect structure relative to non-topological theories with identical 0-form symmetry.  Let us begin once again with $\Z_2$, in the form of SPT$(\Z_2)$, the field theory of a trivially-acting $\Z_2$ symmetry.  One can regard the symmetry of such a theory as $\Z_2^0.\Z_2^1$, similar but opposite to a 2-group.  Since this theory includes a gaugeable 0-form $\Z_2$ symmetry, we would expect to find gauge defects fusing like 2Rep$(\Z_2)$.

In actually gauging that symmetry, we produce two copies of the trivial theory (which can be written as an SPT for the trivial group).  That is,
\be
\label{z2_gauge_theory}
\text{SPT}(\Z_2)/\Z_2^0 = \text{SPT}(\id)\oplus\text{SPT}(\id).
\ee
The result clearly has a $\Z_2^1$ symmetry, which we can gauge to get (for either choice of discrete torsion) a single copy of the trivial theory.  These gaugings can be composed to get
\be
\text{SPT}(\Z_2)/(\Z_2^0.\Z_2^1) = \text{SPT}(\id).
\ee
From this we see that the $\Z_2^0$ gauge defect is no longer simple, as we have 
\be
\label{z2_spt_defects}
\D(\Z_2^0)=\D(\Z_2^0.\Z_2^1)\oplus\D(\Z_2^0.\Z_2^1).
\ee
The fusion of the simple objects is then
\be
\begin{tabular}{r c | c c }
	& $\otimes$ & 1 & $\alpha$ \\\cline{2-4}
	$\D(\id)=$ & 1 & 1 & $\alpha$ \\
	$\D(\Z_2^0.\Z_2^1)=$ & $\alpha$ & $\alpha$ & $\alpha$
\end{tabular}.
\ee

As a quick consistency check, combining the above fusion rules with (\ref{z2_spt_defects}) tells us that $\D(\Z_2^0)$ fuses like 2Rep$(\Z_2)$, as we would expect.

We can perform a similar exercise for SPT$(\Z_2\times\Z_2)$, which experiences the same phenomenon.  The $\Z_2^0\times\Z_2^0$ gauge defects are no longer simple.  Gauging any of the $\Z_2$ subgroups of SPT$(\Z_2\times\Z_2)$ produces 2SPT$(\Z_2)$, which can be reduced to a single copy of SPT$(\Z_2)$ by gauging the 1-form symmetry.  Similarly, gauging the full $\Z_2^0\times\Z_2^0$ without discrete torsion produces four copies of SPT$(\id)$, making it equivalent to four copies of $\D(\Z_2^0\times\Z_2^0).(\Z_2^1\times\Z_2^1)$.  Gauging $\Z_2^0\times\Z_2^0$ with non-trivial discrete torsion simply returns the original theory, so its associated gauge defect is the trivial one.  This leaves us with five simple gauge defects with fusion rules

\be
	\begin{tabular}{r c | c c c c c}
		& $\otimes$ & 1 & $\alpha$ & $\beta$ & $\gamma$ & $\delta$ \\\cline{2-7}
		$\D(\id)=$ & 1 & 1 & $\alpha$ & $\beta$ & $\gamma$ & $\delta$ \\
		$\D(\Z_2^0.\Z_2^1)=$ & $\alpha$ & $\alpha$ & $\alpha$ & $\delta$ & $\delta$ & $\delta$ \\
		$\D(\Z_2^0.\Z_2^1)=$ & $\beta$ & $\beta$ & $\delta$ & $\beta$ & $\delta$ & $\delta$ \\
		$\D(\Z_2^0.\Z_2^1)=$ & $\gamma$ & $\gamma$ & $\delta$ & $\delta$ & $\gamma$ & $\delta$ \\
		$\D((\Z_2^0\times\Z_2^0).(\Z_2^1\times\Z_2^1))=$ & $\delta$ & $\delta$ & $\delta$ & $\delta$ & $\delta$ & $\delta$ \\
	\end{tabular},
\ee
where one can once again check that these results are consistent with the 0-form gauge defects fusing as 2Rep$(\Z_2\times\Z_2)$.\\

A general 2d TQFT takes the form of a direct sum of SPTs.  When some of those copies are isomorphic, there will be a 0-form symmetry given by interchanging them.  For instance $\Z_2$ gauge theory (\ref{z2_gauge_theory}) has a $\Z_2^0$ exchange symmetry, which is the quantum dual to the $\Z_2^0$ by which we gauged SPT$(\Z_2)$.  This 0-form exchange symmetry has a mixed anomaly with the $\Z_2^1$ arising from the direct sum -- since the exchange symmetry swaps vacua, we cannot consistently project onto one universe while also gauging the exchange.  This puts us in the situation of section~\ref{sec:anomalies}, and the intuition developed there should apply to the present example.

So we expect three gauge defects: two from gauging $\Z_2^1$ with and without discrete torsion and one from gauging the exchange $\Z_2^0$.  The $\Z_2^1$ defects $\D(\Z_2^1)_\pm$ are built from condensing the projectors $\pi_{\pm}$ and therefore have orthonormal fusion.  The defect $\D(\Z_2^0)$ arising from the exchange symmetry does not break into a piece on each universe, so we do not find an overall decomposition of the gauge defect fusion.

In order to calculate $\D(\Z_2^0)\otimes\D(\Z_2^1)_\pm$, we note that upon gauging the exchange symmetry, the two vacua recombine into topological local operators $\sigma_1$, $\sigma_g$ which fuse like $\Z_2$, except the $\sigma_g$ are now restricted to the endpoints of $\Z_2$ TDLs implementing the trivially-acting symmetry in the resulting SPT$(\Z_2)$.  Thus, gauging $\pi_pm$ in the resulting theory requires also insertions of the $\hat{\Z}_2^0$ lines, which gauges the SPT, returning us to $\Z_2$ gauge theory.  As with the mixed anomaly example of section~\ref{sec:anomalies}, we find the non-commutative fusion $\D(\Z_2^0)\otimes\D(\Z_2^1)_\pm=D(\Z_2^1)$.  A similar process governs the other direction $\D(\Z_2^1)_\pm\otimes\D(\Z_2^0)$, in which the $\Z_2^0$ exchange lines end up bound to the $-1$-form symmetry dual to $\Z_2^1$, so the only way to gauge $\Z_2^0$ in a single universe would be to first gauge the $\hat{\Z}_2^{-1}$, restoring the direct sum.  In total, then, we find the fusion rules

\be
	\begin{tabular}{r c | c c c}
		& $\otimes$ & $1_+$ & $1_-$ & $\alpha$ \\\cline{2-5}
		$\D(\Z_2^1)_+=$ & $1_+$ & $1_+$ & 0 & $\alpha$ \\
		$\D(\Z_2^1)_-=$ & $1_-$ & $0$ & $1_-$ & $\alpha$ \\
		$\D(\Z_2^0)=$ & $\alpha$ & $1_+$ & $1_-$ & $2\alpha$ \\
	\end{tabular}.
\ee

\section{Discussion}

We have given a qualitatively diverse handful of examples of the fusion of gauge defects in 1d and 2d theories.  Naturally the situation will become more complicated in higher dimensions.  Let us briefly contemplate what to expect from (2+1)-dimensional theories.  Consider a 3d theory with 0-form symmetry $G^0$.  Gauging this produces a theory with the quantum symmetry Rep$(G)^1$.  We expect the 3d gauge defects for this 1-form symmetry to fuse as 3Rep$(G)$.  Gauge defects for the 0-form symmetries themselves will likewise be described by 3-categories.  The interpretation of the self-fusion of these gauge defects will, in line with the examples we have seen in lower dimensions, involve gauging trivially-acting symmetries in 3d.  This will produce TQFT partition functions as the coefficients.  Topological theories in 3d, however, are much richer than those in 2d, as they no longer simply decompose into a direct sum of SPT phases.  So the associated higher categories are more complicated, and are the subject of ongoing mathematical and physical work (see e.g.~\cite{Choi:2022zal,Bartsch:2023pzl,Bhardwaj:2023ayw} for a few physics examples).

Stepping back from the calculation of specific examples, the viewpoint espoused here is that gauging is encoded in the spectrum of $-1$-form symmetries in a theory.  Regarding the act of gauging a symmetry as a symmetry itself allows us to unify two separate elements of field theory, which is intellectually satisfying if nothing else.

\appendix

\section{Relation to Theta Defects}
\label{thetadefects}

There is another method of obtaining condensation defects which can be applied to gauge defects \cite{UNI}.  Briefly, the idea is: given a theory $T$ with symmetry $G$, take the direct product of $T$ with a $G$-symmetric TQFT.  Gauging the diagonal $G$ symmetry of this product produces $T/G$ with a condensation defect (made from the quantum symmetry) inserted; moreover, the choice of these defects corresponds one-to-one with the choice of TQFT.  Note that $G$ here may be higher-form or non-invertible.  Normally in this construction one is interested in condensation defects of lower dimensionality than $T$, so accordingly the TQFT chosen is a lower-dimensional one.  In the context of gauge defects we will take $T$ and the TQFT to be of the same dimension.

\subsection{$\Z_2$ Gauge Defects from Theta Defects}

A brief example will confirm that this construction also applies to gauge defects.  Let $T$ be a 2d theory with (0-form) $\Z_2$  symmetry.  There are two $\Z_2$-symmetric 2d TQFTs available to us -- the $\Z_2$ SPT and its corresponding symmetry breaking phase, known variously as $\Z_2$ gauge theory or $\Z_2$ Dijkgraaf-Witten theory.  We will denote these theories SPT$(\Z_2)$ and DW$(\Z_2)$, respectively.  We can consider using both of them in the theta defect construction to insert gauge defects into $T/\Z_2$.

First we consider the SPT.  The symmetry in such a theory is trivially-acting, which means that adding it will not change $T$ as a local theory.  This is readily verified by calculating the partition function of $[T\otimes \text{SPT}(\Z_2)]/\Z_2$.  Since we are taking the diagonal symmetry, each sector is simply the product of the corresponding sector we would obtain by gauging the theories separately.  The SPT, by definition, has unit partition function on any surface and with any TDL insertions, so the partition function is the same one we would obtain were the SPT not present (entirely in line with the view of it as a trivially-acting symmetry).  Thus,
\be
\label{td_spt}
[T\otimes \text{SPT}(\Z_2)]/\Z_2 = T/\Z_2.
\ee

Now we turn to DW$(\Z_2)$.  One way to obtain this theory is
\be
\text{DW}(\Z_2)=\text{SPT}(\Z_2)/\Z_2=\text{SPT}(\id)\oplus\text{SPT}(\id),
\ee
from which we see that the theory has two ground states, and its $\Z_2$ symmetry is given by their exchange.  Gauging this symmetry should return us to the SPT with a single ground state, and this is able to happen because there are no states fixed under this exchange symmetry (it is freely acting).  This means that every sector other than the trivial one (i.e. the parent theory) is empty.  When gauging $T\otimes\text{DW}(\Z_2)$, then, we find that\footnote{If we are being extra careful about tracking operators, the right-hand side of (\ref{td_spt}) and (\ref{td_dw}) should both be multiplied by SPT$(\Z_2)$.  But this is fine to ignore, as it does not affect the local theory we obtain.}
\be
\label{td_dw}
[T\otimes\text{DW}(\Z_2)]/\Z_2=T.
\ee
As promised, we obtain different results based on the TQFT chosen.  Adding SPT$(\Z_2)$ did not change the result of the gauging, so we can identify it as corresponding to the trivial gauge defect $\D(\id)$.  DW$(\Z_2)$, on the other hand, caused us to obtain $T$ instead of $T/\Z_2$, meaning that including it in the gauging had the same effect as inserting $\D(\hat{\Z}_2)$, the non-trivial gauge defect for the $\hat{\Z}_2$ dual to $\Z_2$.  As promised, the two possible TQFTs generate the two gauge defects that exist in $T/\Z_2$.

We can further use this method to obtain fusion rules of gauge defects.  $\D(\hat{\Z}_2)\otimes\D(\hat{\Z}_2)$ should correspond to gauging $T\otimes\text{DW}(\Z_2)\otimes\text{DW}(\Z_2)$.  This is equivalent taking four copies of $T$ and gauging a $\Z_2$ that both acts on $T$ and exchanges two pairs of copies.  Once again the non-trivial sectors are all empty, but there are now four vacua fixed under the trivial action.  That is,
\be
[T\otimes\text{DW}(\Z_2)\otimes\text{DW}(\Z_2)]/\Z_2 = [4T]/\Z_2 = 2T
\ee
from which we could infer the fusion rule $\D(\hat{\Z}_2)\otimes\D(\hat{\Z}_2)=2\D(\hat{\Z}_2)$.

\subsection{Rep$(S_3)$ Gauge Defects from Theta Defects}
\label{sec:rs3_defects}

We can use this construction to confirm the fusion rules (\ref{2reps3}) for gauge defects in a Rep$(S_3)$-symmetric theory.  There are four minimal $S_3$-symmetric TQFTs from which we can choose, which are helpfully reviewed in \cite[section 4.6]{bhardwaj2023gapped}.  For our purposes it will be most helpful to write the theories as direct sums of SPTs and specify the action of $S_3$ on them.  These theories are:
\begin{itemize}
	\item SPT$(S_3)$, in which all of $S_3$ is trivially-acting.
	\item $\text{SPT}(S_3)/\Z_3=2\text{SPT}(\Z_3)$, in which the $\Z_3$ subgroup of $S_3$ acts trivially and the $\Z_2$ swaps the two copies of SPT$(\Z_3)$.
	\item $\text{SPT}(\text{Rep}(S_3))/(1+X)=3\text{SPT}(\Z_2)$, in which $\Z_3$ acts freely to exchange the three copies and each of the three $\Z_2$ subgroups of $S_3$ swaps two of the copies while leaving the third fixed.
	\item $\text{SPT}(\text{Rep}(S_3))/(1+X+2Y)=6\text{SPT}(\id)$, six copies of the trivial theory with a free action of $S_3$.
\end{itemize}
In order to compute fusion rules for Rep$(S_3)$ gauge defects, we take a theory $T$ with $S_3$ symmetry and will compute the genus one partition function obtained when gauging the direct product of $T$ with the TQFTs above.  Letting $a^2=b^3=1$ with $aba=b^2$ generate $S_3$, an $S_3$ orbifold partition function takes the general form
\begin{multline}
\frac{1}{6}\bigg[({\scriptstyle 1}\square_1)+({\scriptstyle 1}\square_a+{\scriptstyle a}\square_1+{\scriptstyle a}\square_a)+({\scriptstyle 1}\square_{ab}+{\scriptstyle ab}\square_1+{\scriptstyle ab}\square_{ab})+({\scriptstyle 1}\square_{ab^2}+{\scriptstyle ab^2}\square_1+{\scriptstyle ab^2}\square_{ab^2})\\
+({\scriptstyle 1}\square_b+{\scriptstyle 1}\square_{b^2}+{\scriptstyle b}\square_1+{\scriptstyle b^2}\square_1+{\scriptstyle b}\square_b+{\scriptstyle b^2}\square_{b^2}+{\scriptstyle b}\square_{b^2}+{\scriptstyle b^2}\square_b)\bigg].
\label{s3_orb}
\end{multline}
When we add in a TQFT and gauge the diagonal symmetry, each sector in (\ref{s3_orb}) gets multiplied by the partition function of the TQFT in that sector, which is simply the number of vacua fixed by the TDLs wrapping the torus in that sector.  We can then easily determine the effect of inserting each of the four TQFTs:
\begin{multline}
[T\otimes\text{SPT}(S_3)]/S_3=\\
\frac{1}{6}\bigg[1({\scriptstyle 1}\square_1)+1({\scriptstyle 1}\square_a+{\scriptstyle a}\square_1+{\scriptstyle a}\square_a)+1({\scriptstyle 1}\square_{ab}+{\scriptstyle ab}\square_1+{\scriptstyle ab}\square_{ab})+1({\scriptstyle 1}\square_{ab^2}+{\scriptstyle ab^2}\square_1+{\scriptstyle ab^2}\square_{ab^2})\\
+1({\scriptstyle 1}\square_b+{\scriptstyle 1}\square_{b^2}+{\scriptstyle b}\square_1+{\scriptstyle b^2}\square_1+{\scriptstyle b}\square_b+{\scriptstyle b^2}\square_{b^2}+{\scriptstyle b}\square_{b^2}+{\scriptstyle b^2}\square_b)\bigg]=T/S_3,
\end{multline}
\begin{multline}
[T\otimes 2\text{SPT}(\Z_3)]/S_3=\\
\frac{1}{6}\bigg[2({\scriptstyle 1}\square_1)+0({\scriptstyle 1}\square_a+{\scriptstyle a}\square_1+{\scriptstyle a}\square_a)+0({\scriptstyle 1}\square_{ab}+{\scriptstyle ab}\square_1+{\scriptstyle ab}\square_{ab})+0({\scriptstyle 1}\square_{ab^2}+{\scriptstyle ab^2}\square_1+{\scriptstyle ab^2}\square_{ab^2})\\
+2({\scriptstyle 1}\square_b+{\scriptstyle 1}\square_{b^2}+{\scriptstyle b}\square_1+{\scriptstyle b^2}\square_1+{\scriptstyle b}\square_b+{\scriptstyle b^2}\square_{b^2}+{\scriptstyle b}\square_{b^2}+{\scriptstyle b^2}\square_b)\bigg]=T/\Z_3,
\end{multline}
\begin{multline}
[T\otimes 3\text{SPT}(\Z_2)]/S_3=\\
\frac{1}{6}\bigg[3({\scriptstyle 1}\square_1)+1({\scriptstyle 1}\square_a+{\scriptstyle a}\square_1+{\scriptstyle a}\square_a)+1({\scriptstyle 1}\square_{ab}+{\scriptstyle ab}\square_1+{\scriptstyle ab}\square_{ab})+1({\scriptstyle 1}\square_{ab^2}+{\scriptstyle ab^2}\square_1+{\scriptstyle ab^2}\square_{ab^2})\\
+0({\scriptstyle 1}\square_b+{\scriptstyle 1}\square_{b^2}+{\scriptstyle b}\square_1+{\scriptstyle b^2}\square_1+{\scriptstyle b}\square_b+{\scriptstyle b^2}\square_{b^2}+{\scriptstyle b}\square_{b^2}+{\scriptstyle b^2}\square_b)\bigg]=T/\Z_2,
\end{multline}
\begin{multline}
[T\otimes 6\text{SPT}(\id)]/S_3=\\
\frac{1}{6}\bigg[6({\scriptstyle 1}\square_1)+0({\scriptstyle 1}\square_a+{\scriptstyle a}\square_1+{\scriptstyle a}\square_a)+0({\scriptstyle 1}\square_{ab}+{\scriptstyle ab}\square_1+{\scriptstyle ab}\square_{ab})+0({\scriptstyle 1}\square_{ab^2}+{\scriptstyle ab^2}\square_1+{\scriptstyle ab^2}\square_{ab^2})\\
+0({\scriptstyle 1}\square_b+{\scriptstyle 1}\square_{b^2}+{\scriptstyle b}\square_1+{\scriptstyle b^2}\square_1+{\scriptstyle b}\square_b+{\scriptstyle b^2}\square_{b^2}+{\scriptstyle b}\square_{b^2}+{\scriptstyle b^2}\square_b)\bigg]=T,
\end{multline}
from which we see that the addition of SPT$(S_3)$ corresponds to inserting $\D(1)$ into $T/S_3$, 2SPT$(\Z_3)$ to $\D(1+X)$, 3SPT$(\Z_2)$ to $\D(1+Y)$ and 6SPT$(\id)$ to $\D(1+X+2Y)$.  Finding the fusion of these gauge defects is then straightforward.  For example,
\begin{multline}
[T\otimes 3\text{SPT}(\Z_2)\otimes 2\text{SPT}(\Z_3)]/S_3=\\
\frac{1}{6}\bigg[3\cdot 2({\scriptstyle 1}\square_1)+1\cdot0({\scriptstyle 1}\square_a+{\scriptstyle a}\square_1+{\scriptstyle a}\square_a)+1\cdot0({\scriptstyle 1}\square_{ab}+{\scriptstyle ab}\square_1+{\scriptstyle ab}\square_{ab})+1\cdot 0({\scriptstyle 1}\square_{ab^2}+{\scriptstyle ab^2}\square_1+{\scriptstyle ab^2}\square_{ab^2})\\
+0\cdot 2({\scriptstyle 1}\square_b+{\scriptstyle 1}\square_{b^2}+{\scriptstyle b}\square_1+{\scriptstyle b^2}\square_1+{\scriptstyle b}\square_b+{\scriptstyle b^2}\square_{b^2}+{\scriptstyle b}\square_{b^2}+{\scriptstyle b^2}\square_b)\bigg]=T
\end{multline}
which tells us that $\D(1+Y)\otimes\D(1+X)=\D(1+X+2Y)$, and
\begin{multline}
[T\otimes 3\text{SPT}(\Z_2)\otimes 3\text{SPT}(\Z_2)]/S_3=\\
\frac{1}{6}\bigg[3\cdot 3({\scriptstyle 1}\square_1)+1\cdot 1({\scriptstyle 1}\square_a+{\scriptstyle a}\square_1+{\scriptstyle a}\square_a)+1\cdot 1({\scriptstyle 1}\square_{ab}+{\scriptstyle ab}\square_1+{\scriptstyle ab}\square_{ab})+1\cdot 1({\scriptstyle 1}\square_{ab^2}+{\scriptstyle ab^2}\square_1+{\scriptstyle ab^2}\square_{ab^2})\\
+0\cdot 0({\scriptstyle 1}\square_b+{\scriptstyle 1}\square_{b^2}+{\scriptstyle b}\square_1+{\scriptstyle b^2}\square_1+{\scriptstyle b}\square_b+{\scriptstyle b^2}\square_{b^2}+{\scriptstyle b}\square_{b^2}+{\scriptstyle b^2}\square_b)\bigg]=T\oplus T/\Z_2
\end{multline}
which tells us that $\D(1+Y)\otimes\D(1+Y)=\D(1+Y)\oplus\D(1+X+2Y)$.  In total we recover (\ref{2reps3}) with this method.

\subsection{$S_3$ Gauge Defects from Theta Defects}
\label{sec:s3_defects}

The natural next step would be to turn the discussion around and derive gauge defects for an $S_3$-symmetric theory from TQFTs with Rep$(S_3)$ symmetry.  Here, though, we run into a new issue -- in a theory with Rep$(S_3)\times$Rep$(S_3)$ symmetry, there is no diagonal Rep$(S_3)$ subsymmetry.  It is easy to see that this is the case -- writing the simple objects of Rep$(S_3)\times$Rep($S_3$) as pairs $(L_1,L_2)$ of simple objects in Rep$(S_3)$, we see that the diagonal element $(Y,Y)$ squares to
\begin{multline}
(Y,Y)\otimes(Y,Y)=(1+X+Y,1+X+Y)\\
=(1,1)+(1,X)+(1,Y)+(X,1)+(X,X)+(X,Y)+(Y,1)+(Y,X)+(Y,Y)
\end{multline}
which includes non-diagonal elements.  Another issue is that the weight of $(Y,Y)$ is 4 rather than 2.  So the theta defect prescription of taking the direct product of two $G$-symmetric theories and gauging the diagonal $G$ subgroup would seem to run into trouble if we want to allow non-invertible symmetries.

However, there should still be a way to salvage the method.  Recall that, when gauging a non-invertible symmetry, the simple objects appearing in the algebra object need not form a subcategory under fusion -- e.g.~gauging $1+Y$ in Rep$(S_3)$.  Noting that Rep$(S_3)\times$Rep$(S_3)$ is (non-canonically)\cite{mo_groupprod} isomorphic to Rep$(S_3\times S_3)$, and using the fact that gaugings of Rep$(G)$ are in one-to-one correspondence with those of $G$, there ought to be a gauging of Rep$(S_3)\times$Rep$(S_3)$ which corresponds to gauging the diagonal subgroup in $S_3\times S_3$.  To make this more precise, let $A$ and $B$ be 2d theories with $S_3$ symmetry.  There should exist a special symmetric Frobenius algebra $\A$ in Rep$(S_3)\times$Rep$(S_3)$ such that
\be
\label{fc_diag}
\frac{A/S_3\otimes B/S_3}{\A}=\frac{A\otimes B}{S_3}
\ee
where the $S_3$ action on the right-hand side is the diagonal one.\\

We can check whether this setup leads to the expected $S_3$ gauge defects.  Doing so will require enumerating Rep$(S_3)$-symmetric TQFTs \cite[section 5.3]{bhardwaj2023gapped} and writing them as $S_3$ orbifolds.  These theories are
\begin{itemize}
	\item $\text{SPT}(\text{Rep}(S_3)) = [\text{SPT}(\text{Rep}(S_3))/(1+X+2Y)]/S_3$, the trivial Rep$(S_3)$-symmetric phase.
	\item $\text{SPT}(\text{Rep}(S_3))/(1+Y) = 3\text{SPT}(\Z_2) = [\text{SPT}(\text{Rep}(S_3))/(1+X)]/S_3$, where $X$ acts trivially and $Y$ acts on any given vacuum to produce a sum of the other two vacua.
	\item $\text{SPT}(S_3)/\Z_2=2\text{SPT}(\Z_3)=[\text{SPT}(S_3)/\Z_3]/S_3$, where $X$ swaps the two vacua and $Y$ acts on either of them to produce their sum.
	\item $\text{SPT}(S_3)/S_3 = 2\text{SPT}(\id)\oplus\text{SPT}(\Z_2\times\Z_2)$, where the action of Rep$(S_3)$ on the three vacua is given in \cite[(2.11)-(2.15)]{TopOps} and \cite[section 5.3.4]{bhardwaj2023gapped}.
\end{itemize}

Let $T$ be a theory with $S_3$ symmetry.  Applying (\ref{fc_diag}) and using the results from \ref{sec:rs3_defects} above leads to
\begin{align}
\frac{T/S_3\otimes\text{SPT}(\text{Rep}(S_3))}{\A} =\frac{T\otimes 6\text{SPT}(\id)}{S_3}&=T\\
\frac{T/S_3\otimes\text{SPT}(\text{Rep}(S_3))/(1+Y)}{\A} =\frac{T\otimes\text{SPT}(\text{Rep}(S_3))/(1+X)}{S_3}&=T/\Z_2\\
\frac{T/S_3\otimes\text{SPT}(S_3)/\Z_2}{\A} =\frac{T\otimes\text{SPT}(S_3)/\Z_3}{S_3}&=T/\Z_3\\
\frac{T/S_3\otimes\text{SPT}(S_3)/S_3}{\A}=\frac{T\otimes\text{SPT}(S_3)}{S_3}&=T/S_3
\end{align}
and we see that the four Rep$(S_3)$-symmetric minimal TQFTs are in one-to-one correspondence with the four simple $S_3$ gauge defects, as we would expect.  In particular the SPT phase corresponds to the trivial gauge defect and the total SSB phase corresponds to gauging the entire symmetry.

\addcontentsline{toc}{section}{References}

\bibliographystyle{utphys}
\bibliography{Gauging}

\providecommand{\href}[2]{#2}\begingroup\raggedright\begin{thebibliography}{10}

\bibitem{GKSW}
D.~Gaiotto, A.~Kapustin, N.~Seiberg, and B.~Willett, ``{Generalized Global
  Symmetries},'' \href{http://dx.doi.org/10.1007/JHEP02(2015)172}{{\em JHEP}
  {\bfseries 02} (2015) 172}, \href{http://arxiv.org/abs/1412.5148}{{\ttfamily
  arXiv:1412.5148 [hep-th]}}.

\bibitem{Gaiotto:2019xmp}
D.~Gaiotto and T.~Johnson-Freyd, ``{Condensations in higher categories},''
  \href{http://arxiv.org/abs/1905.09566}{{\ttfamily arXiv:1905.09566
  [math.CT]}}.

\bibitem{RSS}
K.~Roumpedakis, S.~Seifnashri, and S.-H. Shao, ``{Higher Gauging and
  Non-invertible Condensation Defects},''
  \href{http://dx.doi.org/10.1007/s00220-023-04706-9}{{\em Commun. Math. Phys.}
  {\bfseries 401} no.~3, (2023) 3043--3107},
  \href{http://arxiv.org/abs/2204.02407}{{\ttfamily arXiv:2204.02407
  [hep-th]}}.

\bibitem{Choi:2022zal}
Y.~Choi, C.~Cordova, P.-S. Hsin, H.~T. Lam, and S.-H. Shao, ``{Non-invertible
  Condensation, Duality, and Triality Defects in 3+1 Dimensions},''
  \href{http://dx.doi.org/10.1007/s00220-023-04727-4}{{\em Commun. Math. Phys.}
  {\bfseries 402} no.~1, (2023) 489--542},
  \href{http://arxiv.org/abs/2204.09025}{{\ttfamily arXiv:2204.09025
  [hep-th]}}.

\bibitem{UNI}
L.~Bhardwaj, S.~Schafer-Nameki, and J.~Wu, ``{Universal Non-Invertible
  Symmetries},'' \href{http://dx.doi.org/10.1002/prop.202200143}{{\em Fortsch.
  Phys.} {\bfseries 70} no.~11, (2022) 2200143},
  \href{http://arxiv.org/abs/2208.05973}{{\ttfamily arXiv:2208.05973
  [hep-th]}}.

\bibitem{Bartsch:2022mpm}
T.~Bartsch, M.~Bullimore, A.~E.~V. Ferrari, and J.~Pearson, ``{Non-invertible
  Symmetries and Higher Representation Theory I},''
  \href{http://arxiv.org/abs/2208.05993}{{\ttfamily arXiv:2208.05993
  [hep-th]}}.

\bibitem{Bhardwaj:2022kot}
L.~Bhardwaj, S.~Schafer-Nameki, and A.~Tiwari, ``{Unifying Constructions of
  Non-Invertible Symmetries},''
  \href{http://dx.doi.org/10.21468/SciPostPhys.15.3.122}{{\em SciPost Phys.}
  {\bfseries 15} (2023) 122}, \href{http://arxiv.org/abs/2212.06159}{{\ttfamily
  arXiv:2212.06159 [hep-th]}}.

\bibitem{Bartsch:2022ytj}
T.~Bartsch, M.~Bullimore, A.~E.~V. Ferrari, and J.~Pearson, ``{Non-invertible
  Symmetries and Higher Representation Theory II},''
  \href{http://arxiv.org/abs/2212.07393}{{\ttfamily arXiv:2212.07393
  [hep-th]}}.

\bibitem{Delcamp:2023kew}
C.~Delcamp and A.~Tiwari, ``{Higher categorical symmetries and gauging in
  two-dimensional spin systems},''
  \href{http://arxiv.org/abs/2301.01259}{{\ttfamily arXiv:2301.01259
  [hep-th]}}.

\bibitem{Bhardwaj:2023wzd}
L.~Bhardwaj and S.~Schafer-Nameki, ``{Generalized Charges, Part I: Invertible
  Symmetries and Higher Representations},''
  \href{http://arxiv.org/abs/2304.02660}{{\ttfamily arXiv:2304.02660
  [hep-th]}}.

\bibitem{Bartsch:2023pzl}
T.~Bartsch, M.~Bullimore, and A.~Grigoletto, ``{Higher representations for
  extended operators},'' \href{http://arxiv.org/abs/2304.03789}{{\ttfamily
  arXiv:2304.03789 [hep-th]}}.

\bibitem{Bhardwaj:2023ayw}
L.~Bhardwaj and S.~Schafer-Nameki, ``{Generalized Charges, Part II:
  Non-Invertible Symmetries and the Symmetry TFT},''
  \href{http://arxiv.org/abs/2305.17159}{{\ttfamily arXiv:2305.17159
  [hep-th]}}.

\bibitem{BhardwajTachikawa}
L.~Bhardwaj and Y.~Tachikawa, ``{On finite symmetries and their gauging in two
  dimensions},'' \href{http://dx.doi.org/10.1007/JHEP03(2018)189}{{\em JHEP}
  {\bfseries 03} (2018) 189}, \href{http://arxiv.org/abs/1704.02330}{{\ttfamily
  arXiv:1704.02330 [hep-th]}}.

\bibitem{etingof2009fusion}
P.~Etingof, D.~Nikshych, V.~Ostrik, and with an appendix~by Ehud~Meir, ``Fusion
  categories and homotopy theory,'' 2009.

\bibitem{LRS}
L.~Lin, D.~G. Robbins, and E.~Sharpe, ``{Decomposition, Condensation Defects,
  and Fusion},'' \href{http://dx.doi.org/10.1002/prop.202200130}{{\em Fortsch.
  Phys.} {\bfseries 70} no.~11, (2022) 2200130},
  \href{http://arxiv.org/abs/2208.05982}{{\ttfamily arXiv:2208.05982
  [hep-th]}}.

\bibitem{LFS}
T.~Vandermeulen, ``{Lower-Form Symmetries},''
  \href{http://arxiv.org/abs/2211.04461}{{\ttfamily arXiv:2211.04461
  [hep-th]}}.

\bibitem{douglas2018fusion}
C.~L. Douglas and D.~J. Reutter, ``Fusion 2-categories and a state-sum
  invariant for 4-manifolds,'' 2018.

\bibitem{Decoppet}
T.~D. D{\'{e}}coppet, ``The relative deligne tensor product over pointed
  braided fusion categories,''
  \href{http://dx.doi.org/10.1016/j.jalgebra.2022.12.029}{{\em Journal of
  Algebra} {\bfseries 620} (Apr, 2023) 89--112}.
  \url{https://doi.org/10.1016%2Fj.jalgebra.2022.12.029}.

\bibitem{Tachikawa}
Y.~Tachikawa, ``{On gauging finite subgroups},''
  \href{http://dx.doi.org/10.21468/SciPostPhys.8.1.015}{{\em SciPost Phys.}
  {\bfseries 8} no.~1, (2020) 015},
  \href{http://arxiv.org/abs/1712.09542}{{\ttfamily arXiv:1712.09542
  [hep-th]}}.

\bibitem{bhardwaj2023gapped}
L.~Bhardwaj, L.~E. Bottini, D.~Pajer, and S.~Schafer-Nameki, ``Gapped phases
  with non-invertible symmetries: (1+1)d,'' 2023.

\bibitem{Greenough_2010}
J.~Greenough, ``Monoidal 2-structure of bimodule categories,''
  \href{http://dx.doi.org/10.1016/j.jalgebra.2010.06.018}{{\em Journal of
  Algebra} {\bfseries 324} no.~8, (Oct, 2010) 1818--1859}.
  \url{https://doi.org/10.1016%2Fj.jalgebra.2010.06.018}.

\bibitem{mo_groupprod}
\url{https://mathoverflow.net/questions/18002/decomposition-of-representations-of-a-product-group}.

\bibitem{TopOps}
D.~G. Robbins, E.~Sharpe, and T.~Vandermeulen, ``{Decomposition,
  trivially-acting symmetries, and topological operators},''
  \href{http://dx.doi.org/10.1103/PhysRevD.107.085017}{{\em Phys. Rev. D}
  {\bfseries 107} no.~8, (2023) 085017},
  \href{http://arxiv.org/abs/2211.14332}{{\ttfamily arXiv:2211.14332
  [hep-th]}}.

\end{thebibliography}\endgroup

\end{document}